\documentclass[%
 reprint,
 amsmath,amssymb,
 aps,
]{revtex4-2}

\usepackage{graphicx}% Include figure files
\usepackage{dcolumn}% Align table columns on decimal point
\usepackage{bm}% bold math
\usepackage{multirow}
\usepackage{makecell}
\usepackage{array}
\usepackage{booktabs}
\usepackage{chemformula}
\usepackage[utf8]{inputenc}
\usepackage{textgreek}

\newcolumntype{C}[1]{>{\centering\arraybackslash}p{#1}}

\begin{document}

\preprint{APS/123-QED}

\title{High efficiency superconducting diode effect in a gate-tunable double-loop SQUID}

\author{Wyatt Gibbons\textsuperscript{1,2}}
\author{Teng Zhang\textsuperscript{1,2}}
\author{Kevin Barrow\textsuperscript{1,2}}
\author{Tyler Lindemann\textsuperscript{1,2,6}}
\author{Jukka I. V\"ayrynen\textsuperscript{1,5}}
\author{Michael J. Manfra\textsuperscript{1,2,3,4,5,6}}
\affiliation{%
\vspace{2mm}
    \textsuperscript{1}Department of Physics and Astronomy, Purdue University, West Lafayette, Indiana 47907, USA\\
    \textsuperscript{2}Birck Nanotechnology Center, Purdue University, West Lafayette, Indiana 47907, USA\\
    \textsuperscript{3}Elmore Family School of Electrical and Computer Engineering, Purdue University, West Lafayette, Indiana 47907, USA\\
    \textsuperscript{4}School of Materials Engineering, Purdue University, West Lafayette, Indiana 47907, USA\\
 \textsuperscript{5}Purdue Quantum Science and Engineering Institute, Purdue University, West Lafayette, Indiana 47907, USA\\
 \textsuperscript{6}Microsoft Quantum, West Lafayette, Indiana 47907, USA\\
}%

\date{\today}

\begin{abstract}
In superconducting quantum interference devices (SQUIDs), the superconducting diode effect may be generated by interference of multiple harmonic components in the current-phase relationships (CPRs) of different branches forming SQUID loops. Through the inclusion of two gate-tunable Josephson junctions in series in each interference branch of a double-loop SQUID, we demonstrate independent control over both the harmonic content and the amplitude of three interfering CPRs, facilitating significant improvement in the maximum diode efficiency. Through optimized gate-controlled tuning of individual Josephson energies, diode efficiency exceeding 50\% is demonstrated. Flux-dependent oscillations show quantitative agreement with a simple model of SQUID operation.
\end{abstract}

\maketitle

\section{Introduction}
Semiconducting diodes are essential nonreciprocal circuit elements offering significantly higher resistance to current flow in one direction compared to the other. This property is crucial to various applications in semiconductor electronics, including power rectification, signal processing, and the implementation of digital logic gates. Driven in part by the requirements for scaled quantum computing technologies, there has been an increasing interest in the superconducting diode effect (SDE) ~\cite{Marcus2024,Marcus2025,Nichele2024,Schonenberger2023,Greco2024,Leblanc2024,Baumgartner2022,Ando2020,ThinFilmTheory2022,Hou2023,RashbaTheory2022,Graphene2022,GrapheneTheory2022,vdW2022,vdWTheory2022,TopoJJ2018,TopoJJ2020,FiniteA2022,FiniteB2022,FiniteC2022,Tarucha2023,TaruchaAnomalous,Regensberg2023,Regensberg2024,Regensberg2025,SDEReview2025}. The superconducting diode effect is a phenomenon where the switching current of a superconducting device depends on the direction in which an external current bias is applied. The SDE may facilitate the development of a superconducting counterpart to the diode operating within completely lossless electronic circuits.
\par
One platform where the SDE has been observed is in superconducting quantum interference devices. The efficiency of the SDE can be characterized by the parameter $\eta=\frac{I_c^+-|I_c^-|}{I_c^++|I_c^-|}$, where $I_c^+$ and $I_c^-$ are the switching currents of the device in the positive and negative bias directions, respectively. This diode effect appears due to the interference of different harmonic components in the current-phase relationships of Josephson junctions forming SQUID loops~\cite{Souto2022,SDEReview2025}. To date, the maximum observed diode efficiencies in SQUIDs peaked at $\eta \sim 30\%$ ~\cite{Marcus2024,Marcus2025,Nichele2024,Schonenberger2023,Greco2024,Leblanc2024}.

Josephson junctions (JJs) typically have sinusoidal CPRs~\cite{Josephson1962,JJReview2004}. One method of generating non-sinusoidal CPRs occurs in Josephson junctions with very low disorder, where highly transparent Andreev bound states (ABSs) dictate supercurrent transport through the JJ~\cite{Likharev1979,DellaRocca2007,Ciaccia2024,Nichele2020}. In a JJ with $N$ independent ABS modes, the CPR is given by~\cite{ABS1991}:

\begin{equation}
    \label{eq:eq1}
    I_{ABS}(\phi) = \frac{e \Delta}{2\hbar}\sum_{i=1}^N {\frac{\tau_{i} \sin(\phi)}{\sqrt{1-\tau_{i} \sin^2(\frac{\phi}{2})}}}
\end{equation}
where $\Delta$ is the superconducting gap, $\tau_i$ is the transparency of the $i$\textsuperscript{th} ABS mode, and $\phi$ is the phase difference between the two SC leads of the junction. Eq.~(\ref{eq:eq1}) can be written in the form $I_{ABS}(\phi)=\sum_{n=1}^{\infty}A_n\sin(n\phi)$, with the relative magnitude of the higher harmonic components $(n \neq 1)$ increasing significantly as $\tau_i$ approaches 1.

Theoretical work by Souto {\it et al.}~\cite{Souto2022} reported a maximum achievable $\eta \sim 37.4\%$ in a single-loop SQUID and $\eta \sim 53.8\%$ in a double-loop SQUID, where a single ABS mode determined the CPR of each branch. These efficiencies were only achievable when one of the junctions in the modeled device was perfectly transparent with $\tau\equiv1$. Perfectly transparent JJs are very difficult to achieve in practice, with typical reported values of $\tau$ in real JJs being significantly lower ~\cite{Schonenberger2023,Nichele2020,Leblanc2024}. Another explicit limitation of this approach is that the amplitude and $\tau$ of the CPR of each branch are not independently tunable. Tunability may be added if the number of ABS modes in each JJ is made an adjustable parameter; however, the practical implementation of this type of tuning in real devices appears to be challenging.

Alternatively, it has been demonstrated that connecting two junctions with conventional sinusoidal CPRs in series can also produce non-sinusoidal Josephson elements ~\cite{Marcus2024,Bozkurt2023,Barash2018}. The total energy of such a system is given by $E_{tot}=E_{J1}[1-\cos\phi_1]+E_{J2}[1-\cos\phi_2]$, where $E_{J1}$ and $E_{J2}$ are the Josephson energies of the two sinusoidal junctions and $\phi_1$ and $\phi_2$ are the phase differences across each junction. The Josephson energy $E_J$ can be related to the switching current of each junction by the equation $E_J\equiv\frac{\hbar I_c}{2e}$~\cite{tinkham1996}. By minimizing the total energy of this system, one finds~\cite{JJReview2004,Bozkurt2023}:
\begin{equation}
    \label{eq:eq2}
    I_{2JJ}(\phi,E_{J1},E_{J2}) = \frac{e \sigma}{2\hbar} \frac{\tau_{eff} \sin(\phi)}{\sqrt{1-\tau_{eff} \sin^2(\frac{\phi}{2})}}
\end{equation}
where $\phi=\phi_1+\phi_2$ is the total phase difference across both JJs, $\sigma=E_{J1}+E_{J2}$, $\tau_{eff}=\frac{4\rho}{(1+\rho)^2}$, and $\rho=\frac{E_{J1}}{E_{J2}}$. Note that Eq. (\ref{eq:eq1}) and Eq. (\ref{eq:eq2}) have the same form, with $\tau_{eff}$ becoming an `effective transparency' and $\sigma$ an `effective superconducting gap' of the two JJ system. Here, we explore this approach to generating non-sinusoidal CPRs and improving the efficiency of gate-tunable diodes. 

\begin{figure}
    \includegraphics[scale=0.45]{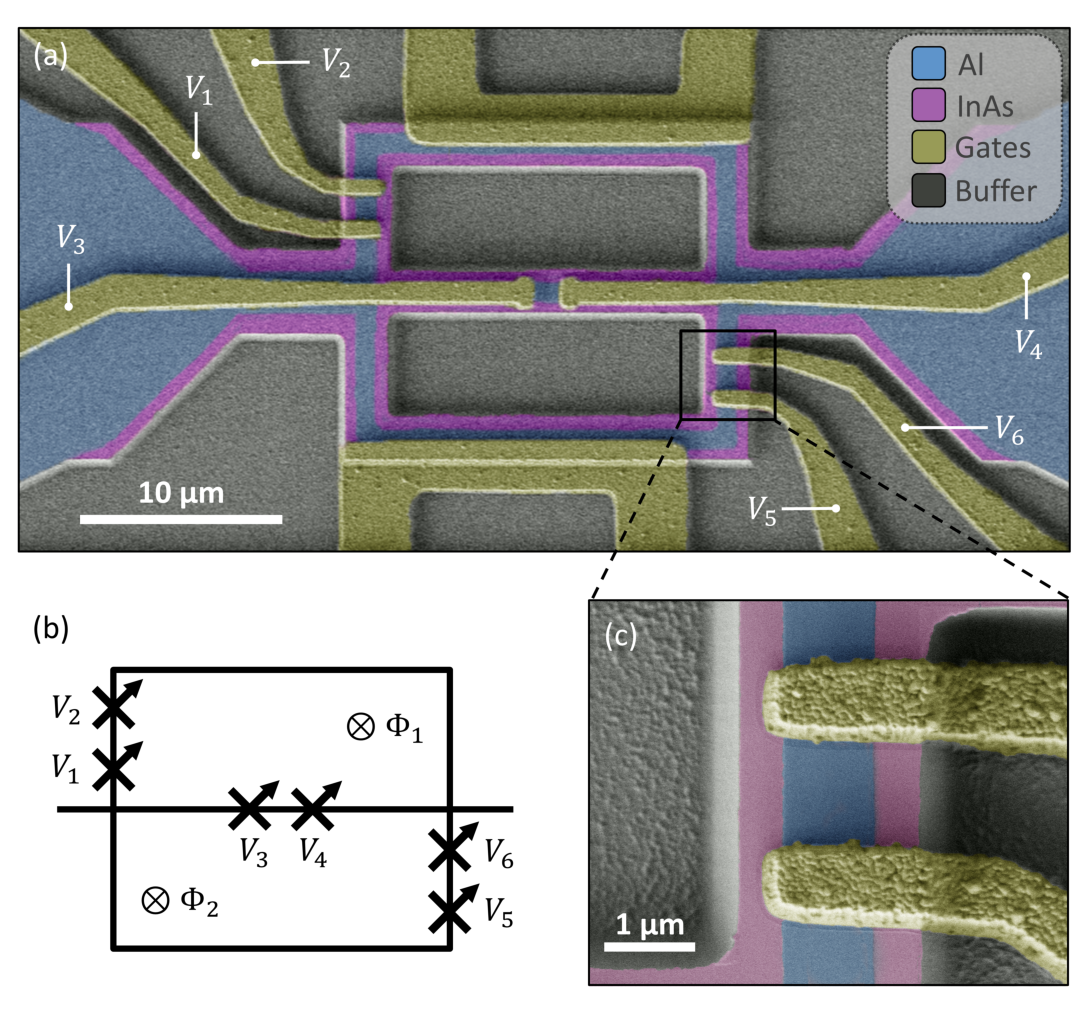}
    \caption{\label{fig:fig1}%
    (a) False-color scanning electron microscope image of a reference device. Two JJs are formed in each of the three branches (blue) of the double-loop SQUID. Electrostatic gates (gold) are then used to tune $E_J$ independently for each junction. Gold loops at the top and bottom of the device may be operated as flux lines to control flux through each loop of the SQUID independently. The black box highlights the region imaged in Fig.~\ref{fig:fig1}(c). (b) Device schematic showing three parallel superconducting branches, each with two gate-tunable JJs in series. The device is flux-biased using an external magnetic field $B_{\perp}$. Both loops of the device are designed to have the same area. (c) False-color scanning electron microscope image of both JJs on the bottom branch of a reference device.}
\end{figure}

Since $\sigma$ and $\tau_{eff}$ in Eq.~(\ref{eq:eq2}) are both dependent on $E_{J1}$ and $E_{J2}$, a device implementing this approach affords significantly more control over the CPR of each branch. It can be shown that in a device where the CPR of each parallel interference branch is given by Eq.~(\ref{eq:eq2}), $\eta=50\%$ and $\eta=63\%$ can be achieved in a single-loop and double-loop SQUID, respectively. Additionally, there is no requirement for highly transparent junctions, as this model assumes each JJ to have a sinusoidal CPR. The only requirement for reaching these efficiencies is that the $E_J$'s within the device can be tuned to precise values. More details on CPRs demonstrating these diode efficiencies are given in Appendices \ref{app:appD} and \ref{app:appE}.

\section{Device CPR and Tuning}
We designed a double-loop SQUID with two JJs in series on each of the three branches of the device, see Fig. 1. The CPR of this double-loop SQUID is given by:
\begin{equation}
    \label{eq:eq3}
    I_{tot}(\phi) = \sum_{i=1}^3 {\frac{e \sigma_i}{2\hbar}\frac{\tau_{eff,i} \sin(\phi+\delta \phi_i)}{\sqrt{1-\tau_{eff,i} \sin^2(\frac{\phi+\delta \phi_i}{2})}}}
\end{equation}
where $\sigma_i=E_{J1,i}+E_{J2,i}$ is the effective SC gap and $\tau_{eff,i}=\frac{4\rho_i}{(1+\rho_i)^2}$ is the effective transparency of the $i$\textsuperscript{th} branch of the device, determined by the ratio $\rho_i=\frac{E_{J1,i}}{E_{J2,i}}$ of the Josephson energies of the two junctions in the $i$\textsuperscript{th} branch. We may choose a global phase such that $\delta \phi_1 = 0$, $\delta \phi_2 = \frac{2\pi\Phi_1}{\Phi_0}$, and $\delta \phi_3 = \frac{2\pi(\Phi_1+\Phi_2)}{\Phi_0}$ with $\Phi_1$ and $\Phi_2$ being the magnetic fluxes piercing each SC loop, as shown in Fig.~\ref{fig:fig1}(b). 
Junctions are indexed according to the convention shown in Fig.~\ref{fig:fig1}(b). Gate voltage $V_i$ is used to tune the coupling of junction $i$, which has a Josephson energy $E_{Ji}$. For instance, the effective transparency and SC gap of the bottom branch of the device are given by $\tau_{eff,3}$ and $\sigma_3$, which are determined by $E_{J5}$ and $E_{J6}$ tuned by gates $V_5$ and $V_6$, respectively. 

\begin{figure}
    \includegraphics[scale=0.5]{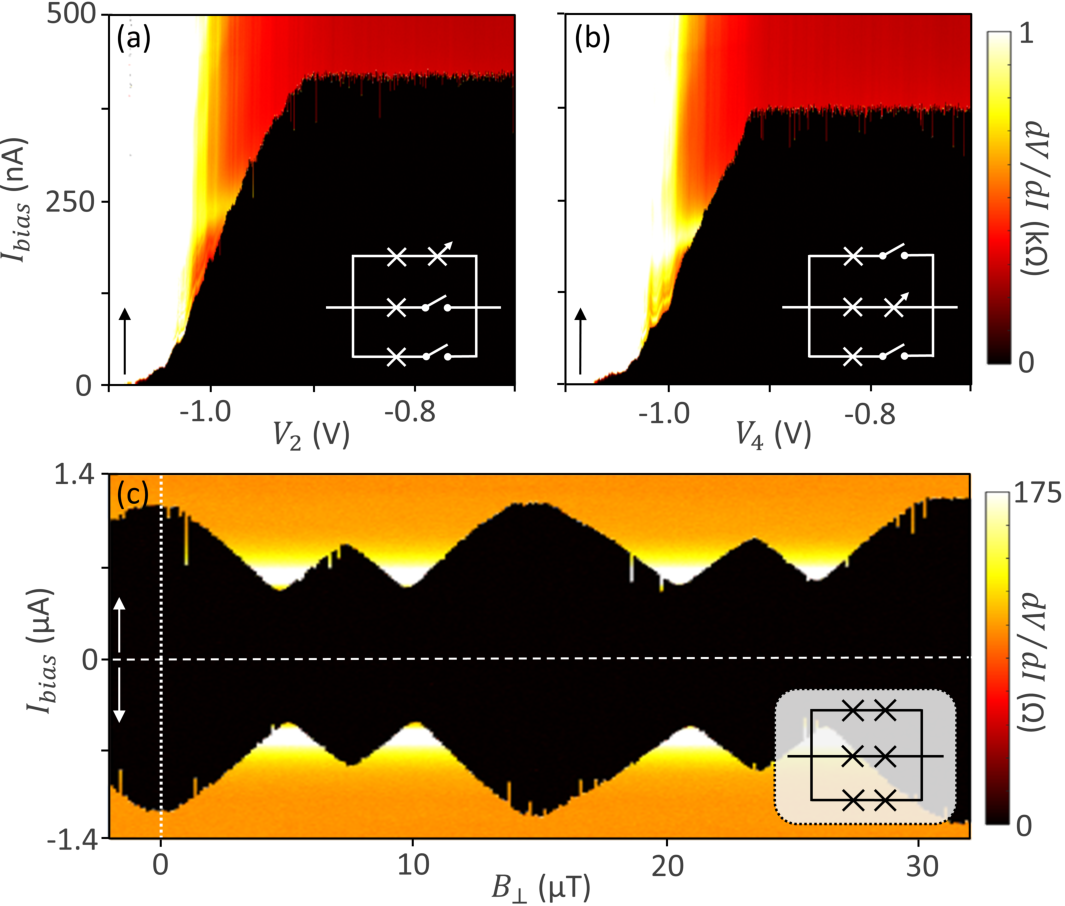}
    \caption{\label{fig:fig2}%
    (a) Differential resistance ($dV/dI$) plotted as a function of gate voltage $V_2$ and DC current bias $I_{bias}$. Gates $V_4$ and $V_6$ are set to $-1.1$V to pinch off the middle and bottom branches, respectively. Black arrow denotes $I_{bias}$ sweep direction for the plotted data. (b) Same as Fig.~\ref{fig:fig2}(a), with top and bottom branches pinched off and gate $V_4$ swept. The color scale applies to both Fig.~\ref{fig:fig2}(a) and Fig.~\ref{fig:fig2}(b). (c) Differential resistance ($dV/dI$) as a function of $I_{bias}$ and perpendicular magnetic field $B_{\perp}$ with all junction gates set to $0$V. White arrows denote $I_{bias}$ sweep directions, with only switching current sweeps plotted. Line cut used to generate IV curve in Fig.~\ref{fig:fig4}(d) denoted by vertical dotted white line.}
\end{figure}

To demonstrate gate control, we first create a quantitative mapping between gate voltage and Josephson energy, $E_{Ji}(V_i)$, for each junction. Fig.~\ref{fig:fig2}(a) shows the $E_J$ vs. gate voltage map for the JJ controlled by gate $V_2$. To generate this data, negative gate voltages were applied to gates $V_4$ and $V_6$ to fully pinch-off the corresponding junctions in the middle and bottom branches of the double-loop SQUID. A gate voltage of -1.1~V proved to be sufficient to pinch off the junctions. We note that gates $V_1$ and $V_3$ were not responsive and held at ground during the course of this study.

With the other branches pinched off, the CPR of the full device is given by Eq.~(\ref{eq:eq2}). From Eq.~(\ref{eq:eq2}), one can show that (see Appendix \ref{app:appC} for further details):

\begin{equation}
    \label{eq:eq4}
       I_c(E_{J1},E_{J2}) = \frac{2e}{\hbar} \textnormal{min}\{E_{J1},E_{J2}\} 
\end{equation}

If we apply a negative gate voltage to $V_2$ to reduce $I_c$ while keeping $V_1=0$ (this occurs at $V_2\sim-0.9$~V in Fig.~\ref{fig:fig2}(a)), $I_c$ is proportional to $E_{J2}$. $V_2$ is then set to a range of voltages between $-0.7$~V and $-1.1$~V. At each voltage, $I_{bias}$ is swept from $0~\textnormal{nA} \rightarrow 500~\textnormal{nA} \rightarrow 0~\textnormal{nA}$ and the voltage drop across the full device is measured to calculate $dV/dI$. In Fig.~\ref{fig:fig2}(a), only the sweeps from $I_{bias}=0~\textnormal{nA} \rightarrow 500~\textnormal{nA}$ are plotted. 

Eq.~(\ref{eq:eq4}) allows us to use the switching current data, $I_c(V_2)$, in Fig.~\ref{fig:fig2}(a) as a `look-up table' to know what voltage $V_2$ we need to set to achieve a specific $E_{J2}$ value. Similar gate voltage vs. $E_J$ maps were created for gates $V_4$ (shown in Fig.~\ref{fig:fig2}(b)), $V_5$, and $V_6$. Maps for $V_5$ and $V_6$ are included in Appendix \ref{app:appF}.

While $V_1$ and $V_3$ are not actively tuned, we still need to be able to approximate $E_{J1}$ and $E_{J3}$ to model the device CPR properly. All junctions are designed to have the same dimensions, with each junction having a designed width of 1~μm and a designed length of 200~nm. For this reason, we make the assumption that $E_{J1}=E_{J2}$ and $E_{J3}=E_{J4}$ when zero external gate voltage is applied. This assumption results in the $E_{J1}$ and $E_{J3}$ values shown in Table \ref{tab:table1}, calculated using the $E_J$ vs. gate voltage maps for $E_{J2}$ and $E_{J4}$. The areas of both SC loops are also designed to be identical. Since flux-biasing was done using an external perpendicular magnetic field $B_{\perp}$, which is uniform over the device area, we assume that $\Phi_1=\Phi_2$. These conditions provide a clear definition of the region of flux space $\{\Phi_1,\Phi_2\}$ explored through flux biasing and allows us to set $\delta \phi_2 = \frac{2\pi\Phi_1}{\Phi_0}$, and $\delta \phi_3 = \frac{4\pi\Phi_1}{\Phi_0}$ when numerically calculating the CPR using Eq.~(\ref{eq:eq3}).

\section{SDE Measurements}
\subsection{All gates at 0~V}
We start by characterizing the device in the configuration where the junctions are nominally identical. Periodic SQUID oscillations in both bias directions with all $V_i=0$~V are displayed in Fig.~\ref{fig:fig2}(c). This data shows largely symmetric periodic oscillations in both $I_c^+$ and $I_c^-$, with a period of approximately 15~μT. For our device, each SQUID loop has a designed area of $A\sim120~$μ$\textnormal{m}^2$ such that we expect switching current oscillations to have a periodicity of $\Phi_0/A\sim17$~μT. The slightly smaller period that we observe may be explained by the fluxoid focusing effect due to the surrounding Al thin film~\cite{FluxFocus2021}.

Switching currents in the positive and negative bias directions are extracted from this data using a threshold method, where the device is said to have switched from a superconducting state to a normal state once the measured $dV/dI$ becomes greater than some threshold $R_{thresh}$. Throughout this report, a threshold resistance of $R_{thresh}=10~\Omega$ is chosen because it is roughly twice the maximum noise amplitude of $dV/dI$ measured in the superconducting state. 

It is worth noting that there is a measurable flux-tunable diode effect revealed by extracting switching currents from the data in Fig.~\ref{fig:fig2}(c), with a maximum magnitude of $|\eta| \sim 6\%$. This asymmetry can be attributed to slight variations in the intrinsic $E_J$'s of each junction. The $E_J$'s of each junction appear to vary by approximately $10-20\%$ at zero bias, which can be demonstrated using gate response data for each junction (see Fig.~\ref{fig:fig2}(a)-(b)). These variations cause the higher harmonics in the CPRs of each branch to have slightly different amplitudes. Once a finite external field is applied, these varying higher harmonic amplitudes can interfere in a way that breaks inversion symmetry and leads to a small but observable diode effect.

\subsection{Voltage configuration yielding $|\eta|>50\%$}
\label{sec:sec3.1}
\begin{figure}
    \includegraphics[scale=0.5]{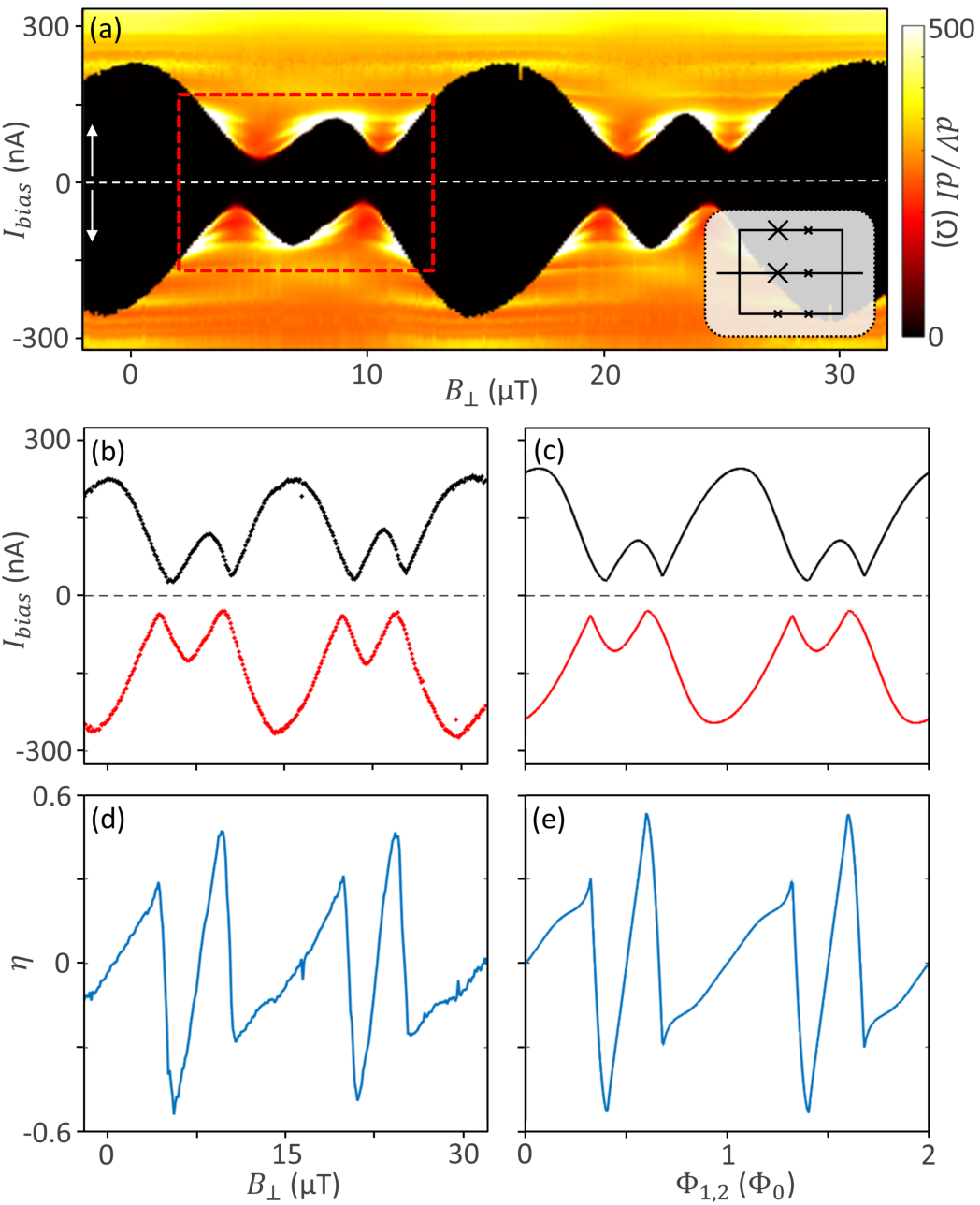}
    \caption{\label{fig:fig3}%
    (a) Differential resistance ($dV/dI$) as a function of $I_{bias}$ and $B_{\perp}$ with junction gates tuned to a configuration to enhance $\eta$ ($V_2=-1.04$~V, $V_4=-1.02$~V, $V_5=-1.013$~V, $V_6=-1.005$~V). Dashed red box denotes $I_{bias}$ and $B_{\perp}$ ranges displayed in Fig.~\ref{fig:fig4}(a). (b) $I_c^+$ (black) and $I_c^-$ (red) as a function of $B_{\perp}$ extracted from the data plotted in Fig.~\ref{fig:fig3}(a) using a threshold resistance of $10~\Omega$. (c) Modeled $I_c^+$ and $I_c^-$ as a function of external magnetic flux threading both loops, where $\Phi_1=\Phi_2$. Calculated from the maximum and minimum values of the device CPR given by Eq.~(\ref{eq:eq3}) using $E_{Ji}$'s displayed in Table \ref{tab:table1}. (d) Measured diode efficiency $\eta$ as a function of $B_{\perp}$ calculated from the switching currents plotted in Fig.~\ref{fig:fig3}(b). (e) $\eta$ as a function of external flux ($\Phi_1=\Phi_2$) calculated from the modeled switching currents in Fig.~\ref{fig:fig3}(c).
    }
\end{figure}

A significant diode efficiency can be achieved by choosing a suitable asymmetric voltage configuration. $E_J$'s necessary to achieve a maximum diode effect are determined using a Monte Carlo optimization method, where the four tunable $E_J$'s and the external flux biases $\Phi_{1,2}$ are sampled from a uniform distribution within the experimentally tunable range and $\eta$ is calculated numerically for each set of device parameters using Eq.~(\ref{eq:eq3}) (see Appendix \ref{app:appE} for more details). Through this optimization method, we determine that $\eta$ is maximized in configurations where the CPR of one branch is maximally non-sinusoidal, with $\tau_{eff}=1$, and the CPRs of the other two branches have significantly less higher harmonic content, with $\tau_{eff} \sim 0.5-0.6$.

This configuration is achieved in our device by setting $E_{J1} \gg E_{J2}$, and $ E_{J3} \gg E_{J4}$, and $E_{J5} \sim E_{J6}$. This leads to the observation of highly asymmetric SQUID oscillations, as is shown in Fig.~\ref{fig:fig3}(a). Switching currents in the positive and negative bias directions are plotted in Fig.~\ref{fig:fig3}(b) for an $R_{thresh}$ of $10~\Omega$.

The minima of $I_c^+$ and $|I_c^-|$ in each period of these SQUID oscillations are clearly shifted from each other, leading to a strong observed diode effect at specific $B_{\perp}$ values. Through flux tuning, we demonstrate the ability to periodically tune $\eta$ to any value between $-54\%$ and $47\%$ in this voltage configuration, see Fig.~\ref{fig:fig3}(d). $\lvert\eta\rvert=54\%$ is a significant increase over previously reported maximum efficiencies reported in the literature ~\cite{Marcus2024,Marcus2025,Nichele2024,Schonenberger2023,Greco2024,Leblanc2024} and demonstrates the potential of our approach.

To model this data, we numerically calculate the device CPR for a range of external flux values from $\Phi_1=\Phi_2\in[0,2\Phi_0]$ using Eq.~(\ref{eq:eq3}) with $E_{Ji}$'s for each junction set to the values shown in Table \ref{tab:table1}. At each $\Phi_{1,2}$, $I_c^+$ and $I_c^-$ are given by the maximum and minimum values of $I_{tot}(\phi)$, respectively. This calculation results in the modeled $I_c^+$ and $I_c^-$ oscillations shown in Fig.~\ref{fig:fig3}(c), where we see broad agreement with our data. We also observe that the measured flux dependence of $\eta$ agrees with this model, as is shown in Fig.~\ref{fig:fig3}(e). We can also demonstrate a similar agreement for a different voltage configuration of this device, where negative gate voltages are applied only to $V_2$ and $V_4$ (see Appendix \ref{app:appF} for data and analysis).

\begin{table}[h]
\caption{\label{tab:table1}
Gate voltages applied to each JJ to generate the data shown in Fig.~\ref{fig:fig3}(a). $E_{J,i}$'s used to calculate modeled $I_c^+$ and $I_c^-$ oscillations in Fig.~\ref{fig:fig3}(c) are also shown, along with the effective transparency of each branch in this model.}
\centering
\begin{tabular*}{\linewidth}{@{\extracolsep{\fill}} c c c c c}
\toprule
$i$ & $V_i$ (V) & $E_{Ji}$ (nA$\cdot\frac{\hbar}{2e}$) & $E_{Ji}$ (meV) & $\tau_{\mathrm{eff}}$ \\
\midrule
1 & 0      & 420 & 0.8633 & 0.538 \\
2 & -1.040 &  80 & 0.1644 &       \\
\midrule
3 & 0      & 380 & 0.7811 & 0.619 \\
4 & -1.020 &  90 & 0.1850 &       \\
\midrule
5 & -1.013 &  80 & 0.1644 & 0.997 \\
6 & -1.005 &  90 & 0.1850 &       \\
\bottomrule
\end{tabular*}
\end{table}

\begin{figure*}
    \includegraphics[scale=0.5]{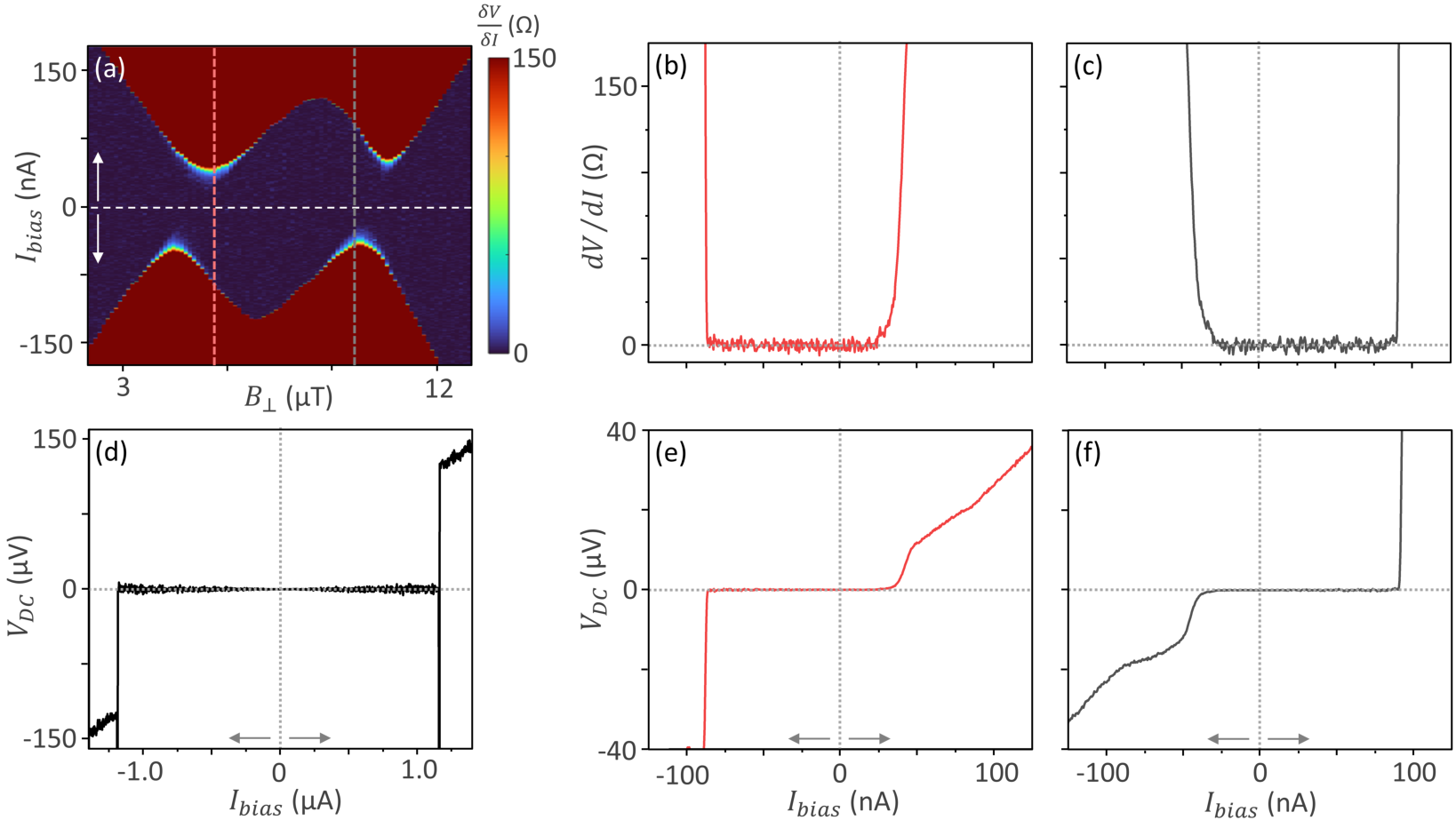}
    \caption{\label{fig:fig4}%
    (a) Differential resistance ($dV/dI$) for the narrow range of $I_{bias}$ and $B_{\perp}$ values denoted by the dotted box in Fig.~\ref{fig:fig3}(a). Note that this is the same data from Fig.~\ref{fig:fig3}(a) plotted using a different colormap to highlight smooth SC transitions. Dotted pink and gray lines denote $dV/dI$ data demonstrating high $|\eta|$ in both polarities, plotted in Fig.~\ref{fig:fig4}(b) and Fig.~\ref{fig:fig4}(c), respectively. (b) $dV/dI$ as a function of $I_{bias}$ at external field $B_{\perp} \sim 5.56$~μT. $\eta = -54\%$ calculated for a threshold resistance of 10~$\Omega$. (c) $dV/dI$ as a function of $I_{bias}$ at external field $B_{\perp}\sim 9.57$~μT. $\eta = 47\%$ calculated for a threshold resistance of 10~$\Omega$. (d) DC IV for device configuration with all junction gates set to $0$~V and external magnetic field set to 0~μT. Arrows at the bottom of the figure indicate $I_{bias}$ sweep direction. (e) DC IV generated from data in Fig.~\ref{fig:fig4}(b). (f) DC IV generated from data in Fig.~\ref{fig:fig4}(c).}
\end{figure*}

We observe an interesting asymmetry in the sharpness of the increase in $dV/dI$ around the SC transition at certain values of $B_{\perp}$ in this voltage configuration. We highlight this asymmetry in Fig.~\ref{fig:fig4}(a), where regions of low but non-zero resistance are clearly present for particular $I_{bias}$ and $B_{\perp}$ values. These observations, indicative of `smoother' superconducting transitions around the field values where we observe the lowest $I_c^+$ and $|I_c^-|$, are in qualitative agreement with basic thermal fluctuation theory~\cite{Ambegaokar1969}. Since the minima of the oscillations in $I_c^+$ and $|I_c^-|$ occur at different $B_{\perp}$, the effect is to smooth the SC transition in one bias direction and generate a sharp SC transition in the other direction (see Fig.~\ref{fig:fig4}(b,c). 

Fig.~\ref{fig:fig4}(b) shows $dV/dI$ measured in both current bias directions at the voltage and flux configuration where $\eta\sim-54\%$ was realized. Here, the asymmetry in the sharpness of the SC transition is apparent, with the increase in $dV/dI$ being significantly more gradual in the positive bias direction. By changing the external flux through the device, we can also observe a mirrored asymmetric transition with the smooth increase in $dV/dI$ in the negative bias direction instead of the positive direction (see Fig.~\ref{fig:fig4}(c)). 

Fig.~\ref{fig:fig4}(d-f) displays a comparison of the DC IV data when all gate voltages are set to $0$~V (Fig.~\ref{fig:fig4}(d)) to the configurations corresponding to the most negative (Fig.~\ref{fig:fig4}(e)) and most positive (Fig.~\ref{fig:fig4}(f)) measured values of $\eta$. In the initial state with the gates set $0$~V, we find $I_c^+ \sim I_c^-$ and $|\eta|<1\%$. We observe a sharp SC transition in both bias directions. Clearly, this initially symmetric current response can be tuned using electrostatic gating and external magnetic flux to operate as an efficient diode.

\section{Conclusions}
In summary, we realize an efficient diode effect in a superconducting interferometer in which non-sinusoidal CPRs are generated by tuning multiple voltage-tunable Josephson junctions. In optimized voltage configurations, we observe highly asymmetric oscillations in $I_c^+$ and $I_c^-$ with respect to external field, signifying a flux-tunable superconducting diode with a maximum $|\eta|>50\%$. These findings demonstrate the ability to engineer the CPR of non-sinusoidal Josephson elements to develop more effective superconducting diodes.

Tunable non-sinusoidal CPRs in superconducting circuits add significant functionality. For example, added control over the CPR and EPR of SQUIDs may allow for specific engineering of SQUID Hamiltonians. There is potential to develop noise-resilient qubits ~\cite{TransmonChargeDisp2020,SCIslandChargeDisp2020,Willsch2024,JJArrayQubit2022,Smith2020}, where the energy levels of the qubit could be controlled by setting specific $E_J$'s to generate a desired device EPR.

\section*{ACKNOWLEDGMENTS}
Funding from Microsoft Quantum is gratefully acknowledged.

\section*{DATA AVAILABILITY}
The data that support the findings of this article may be made available upon request.

\appendix
\section{SAMPLE PREPARATION}
\label{app:appA}
A hybrid superconductor-semiconductor heterostructure is used for device fabrication. This wafer is grown using molecular beam epitaxy (MBE) to ensure a high-quality interface between the superconductor and semiconductor. An InGaAs/InAs/InGaAs layer stack forms a near-surface quantum well that supports a two-dimensional electron gas (2DEG). The superconducting layer is formed by growing 5~nm of Al epitaxially on top of this semiconductor stack.

All device patterning is performed using electron beam lithography (JEOL JBX-8100, 100~keV).  First, the mesa for each device is defined with a wet chemical etch using (220:55:3:3 \ch{H2O}:\ch{C6H8O7}:\ch{H3PO4}:\ch{H2O2}). JJs are then formed by etching the epitaxial Al using Transene Aluminum etch type D at $50^{\circ}$C for 9 seconds. The JJs are designed to have a length of 200~nm and a width of 1~μm. SEM imaging of a reference device measured the length of the junctions after etching to be $\sim230$~nm. A dielectric layer of 18~nm HfOx is then grown using atomic layer deposition (Cambridge Nanotech Fiji). Electrostatic gates are defined by depositing a metal stack of Ti/Al/Ti/Au (15~nm/300~nm/15~nm/100~nm) on top of this dielectric layer using electron beam evaporation (CHA E-beam Evaporator). The thick layer of Al was included in this metal stack because superconducting flux lines are patterned onto our device in the same layer as the electrostatic gates. These flux lines (imaged in Fig.~\ref{fig:fig1}(a)) are not used for this experiment.

Prior to the measurement of the two-loop SQUID used for the main experiment, characterization of this wafer is performed with a gated Hall bar and a superconducting quantum point contact (SQPC) on the same chip. In the gated Hall bar, a peak 2DEG mobility of $\mu_{peak} \sim 65,000\textnormal{ cm}^2/\textnormal{Vs}$ is measured at a charge carrier density of $n \sim 0.67\times10^{12}\textnormal{ cm}^{-2}$. A hard induced superconducting gap of $2\Delta^* \sim 380~$μeV is measured in tunneling spectroscopy measurements performed using the SQPC.

\section{MEASUREMENT SETUP}
\label{app:appB}
Electrical transport measurements are performed in a commercial cryofree dilution refrigerator (Oxford Instruments, Triton 500) at a base temperature of 10~mK. Standard DC and low-frequency AC ($f=101$~Hz) lock-in (Stanford Research, SR860) measurement techniques are used. DC and AC currents are amplified using a commercially available current-to-voltage converter (Basel, SP 963c) with a gain of $10^6$. Commercial voltage preamplifiers (Basel, SP 1004) with a gain of $10^2$ are used to amplify DC and AC voltages. An external magnetic field is applied using a current source meter (Keithley 2612B) connected to the vector magnet of the dilution refrigerator.

All measurement lines are filtered at cryogenic temperatures using RC and RF low-pass filter banks (QDevil) with a cutoff frequency of approximately 50~kHz. Simple RC filters with a cutoff frequency of $\sim 2$~kHz are present on the sample motherboard, adding additional filtering to all lines. Lines for electrostatic gates are additionally filtered at room temperature using homemade RC filters with a cutoff frequency of 10~Hz.

\section{RELATING $I_c$ THROUGH A SINGLE BRANCH TO $E_J$}
\label{app:appC}
Recalling that the relation between Josephson energy and critical current for a single junction is given by $E_J\equiv\frac{\hbar I_c}{2e}$, an equivalent statement to Eq.~(\ref{eq:eq4}) is
\begin{equation}
    \label{eq:2JJ_Ic}
    I_c(E_{J1},E_{J2})=\textnormal{min}\{I_{c1},I_{c2}\}
\end{equation}
where $I_{c1}$ and $I_{c2}$ are the critical currents of each junction. When two JJs are connected in series, the current passing through both junctions must be equivalent since current must be conserved. So, the critical current of a branch with two JJs will be given by the smaller junction critical current, as is stated in Eq.~(\ref{eq:2JJ_Ic}).

We also provide a more rigorous mathematical proof that Eq.~(\ref{eq:eq4}) follows from Eq.~(\ref{eq:eq2}). We first need to express the critical current through a single branch explicitly in terms of only $E_{J1}$ and $E_{J2}$. This can be done by finding an expression for the $\phi$ that maximizes Eq.~(\ref{eq:eq2}). To start, we take the derivative of Eq.~(\ref{eq:eq2}) with respect to $\phi$ and set it equal to zero:
\begin{equation}
    \label{eq:eqS5}
    \frac{\partial I_{2JJ}}{\partial \phi} = \frac{e\sigma\tau}{2^{5/2}\hbar}\left(\frac{\tau \cos(2\phi)-4(\tau-2)\cos(\phi)+3\tau}{[\tau(\cos(\phi)-1)+2]^{3/2}}\right)=0
\end{equation}
where $\sigma=E_{J1}+E_{J2}$, $\tau=\frac{4\rho}{(1+\rho)^2}$, and $\rho=\frac{E_{J1}}{E_{J2}}$. Solving Eq.~(\ref{eq:eqS5}) for $\phi$ yields two real solutions, which we will define as $\phi_+$ and $\phi_-$, where:

\begin{equation}
    \label{eq:eqS6}
    \phi_{\pm}=\pm\cos^{-1}\left(\frac{-2+2\sqrt{1-\tau}+\tau}{\tau}\right)
\end{equation}
$\phi_+$ and $\phi_-$ are the phases corresponding to the maximum and minimum values of Eq.~(\ref{eq:eq2}), respectively. Substituting this expression for $\phi_+$ into Eq.~(\ref{eq:eq2}) allows us to express the critical current through a single branch with two JJs as a function of only $E_{J1}$ and $E_{J2}$:

\begin{equation}
    \label{eq:eqS7}
    I_c(E_{J1},E_{J2})=\frac{e\sigma}{2\hbar}\frac{\tau \sin(\cos^{-1}(\frac{-2+2\sqrt{1-\tau}+\tau}{\tau}))}{\sqrt{1-\tau \sin^2(\frac{1}{2}\cos^{-1}(\frac{-2+2\sqrt{1-\tau}+\tau}{\tau}))}}  
\end{equation}

We can then square both sides of Eq.~(\ref{eq:eqS7}) and expand the product $\sigma \tau$ in terms of $E_{J1}$ and $E_{J2}$ to arrive at the expression:
\begin{equation}
    \label{eq:P1}
    I_c^2=\left[\frac{2eE_{J1}E_{J2}}{\hbar(E_{J1}+E_{J2})}\right]^2\frac{\sin^2\left(\cos^{-1}\left(\frac{-2+2\sqrt{1-\tau}+\tau}{\tau}\right)\right)}{1-\tau \sin^2\left(\frac{1}{2}\cos^{-1}\left(\frac{-2+2\sqrt{1-\tau}+\tau}{\tau}\right)\right)}
\end{equation}
After removing the trigonometric components in the numerator and denominator of this expression using double-angle identities, power-reducing identities, and simplification, Eq.~(\ref{eq:P1}) becomes:
\begin{equation}
    \label{eq:P2}
    I_c^2=\left[\frac{2eE_{J1}E_{J2}}{\hbar(E_{J1}+E_{J2})}\right]^2\frac{1-\left(\frac{-2+2\sqrt{1-\tau}+\tau}{\tau}\right)^2}{\sqrt{1-\tau}}
\end{equation}
Then, taking the square root of both sides and simplifying further:
\begin{equation}
    \label{eq:P3}
    I_c = \frac{4eE_{J1}E_{J2}}{\hbar(E_{J1}+E_{J2})}\sqrt{\frac{2-2\sqrt{1-\tau}-\tau}{\tau^2}}
\end{equation}
After substituting in the expression for $\tau$ in terms of $E_{J1}$ and $E_{J2}$ and simplifying again, we finally arrive at the expression
\begin{equation}
    \label{eq:P4}
    \begin{cases}
        I_c = \frac{2eE_{J1}}{\hbar};\quad & E_{J1} \leq E_{J2} \\

        I_c = \frac{2eE_{J2}}{\hbar};\quad & E_{J1} \geq E_{J2}
    \end{cases}
\end{equation}

Eq.~(\ref{eq:P4}) is an equivalent statement to Eq.~(\ref{eq:eq4}). This proves that the critical current through both junctions is proportional to the smaller of the two $E_J$'s, and allows us to create the gate voltage to $E_J$ maps for each junction.

\section{CPR OPTIMIZING $\eta$ IN A SINGLE-LOOP SQUID}
\label{app:appD}
In a single-loop SQUID, where each interference branch is given by Eq.~(\ref{eq:eq2}), the total CPR is the sum of these two interfering CPRs. For this optimization problem, we temporarily ignore the prefactor $\frac{e}{2\hbar}$ common to each branch and write this CPR as:

\begin{equation}
    \label{eq:eqS1}
    I_{SL}(\phi) \propto \sigma_1\frac{\tau_1 \sin(\phi)}{\sqrt{1-\tau_{1} \sin^2(\frac{\phi}{2})}}+\sigma_2\frac{\tau_2 \sin(\phi+\delta \phi)}{\sqrt{1-\tau_{2} \sin^2(\frac{\phi+\delta \phi}{2})}}
\end{equation}

Note that $\tau_{eff,i}$'s have been rewritten as $\tau_i$'s here for clarity. $I_{SL}(\phi)$ denotes the CPR of a Josephson rhombus, a device measured in previous experiments \cite{Marcus2025}. The initial goal is to find the combination of parameters $\{\sigma_1,\sigma_2,\tau_1,\tau_2,\delta\phi\}$ that would yield a CPR with a maximized $\eta$.

Here, we note a few observations that help to reduce the dimensionality of the parameter space for optimization. First, we note that we expect a maximized $\eta$ when one of the two branches of our device has maximal higher harmonic content. This expectation is motivated by the fact that asymmetry in SQUIDs is caused by the interference of these higher harmonic components. This is also what is observed in Souto's single-ABS model \cite{Souto2022}. For this reason, we set $\tau_1=1$ in Eq.~(\ref{eq:eqS1}) and leave $\tau_2$ as a free parameter.

Also, we can divide Eq.~(\ref{eq:eqS1}) by $\sigma_2$ and define a ratio $\beta=\frac{\sigma_1}{\sigma_2}$ to reduce the number of free parameters further down to three ($\beta$,$\tau_2$, and $\delta\phi$). This allows us to rewrite Eq.~(\ref{eq:eqS1}) as:

\begin{equation}
    \label{eq:eqS2}
    I_{SL}(\phi) \propto \beta\frac{\sin(\phi)}{\sqrt{1- \sin^2(\frac{\phi}{2})}}+\frac{\tau_2 \sin(\phi+\delta \phi)}{\sqrt{1-\tau_{2} \sin^2(\frac{\phi+\delta \phi}{2})}}
\end{equation}

By doing this, the multiplicative prefactor in front of our expression becomes $\frac{e\sigma_2}{2\hbar}$. We can ignore such prefactors for this optimization since they multiply $I_c^+=\textnormal{max}\{I_{SL}(\phi)\}$ and $I_c^-=\textnormal{min}\{I_{SL}(\phi)\}$ by the same value, and therefore have no impact on $\eta$.

We then perform Monte Carlo optimization on Eq.~(\ref{eq:eqS2}), where a large number ($N=10^6$) of $\beta$, $\tau_2$, and $\delta\phi$ values are randomly generated from a uniform distribution within the ranges $\beta \in[0,2], \tau_2 \in[0,1],$ and $\delta\phi\in[0,2\pi]$ and $\eta$ is numerically calculated for each combination. Interestingly, there appears to be convergence towards a single optimal solution. This convergence occurs as $\beta\rightarrow0$, $\tau_2\rightarrow0$, and $\delta\phi\rightarrow\pm\frac{3\pi}{4}$.

Clearly, inserting these values into Eq.~(\ref{eq:eqS2}) yields a trivial result. However, if we divide Eq.~(\ref{eq:eqS2}) by $\tau_2$ and then note that $\lim_{\tau_2\rightarrow0}\frac{\sin(\phi+\delta\phi)}{\sqrt{1-\tau_2 \sin^2(\frac{\phi+\delta\phi}{2})}}=\sin(\phi+\delta\phi)$, we can rewrite this expression again as: 

\begin{equation}
    \label{eq:eqS3}
    I_{SL}(\phi) \propto \frac{\beta}{\tau_2}\frac{\sin(\phi)}{\sqrt{1- \sin^2(\frac{\phi}{2})}}+\sin(\phi+\delta\phi)
\end{equation}

We then perform Monte Carlo optimization again with only two free parameters: $\frac{\beta}{\tau_2}$ and $\delta\phi$. This approach indeed results in convergence towards an optimal set of parameters, where $\frac{\beta}{\tau_2}=\frac{\sqrt{2}}{2} $, $ \delta\phi=\pm\frac{3\pi}{4}$, and $\eta=50\%$.

\begin{figure*}
    \includegraphics[scale=0.5]{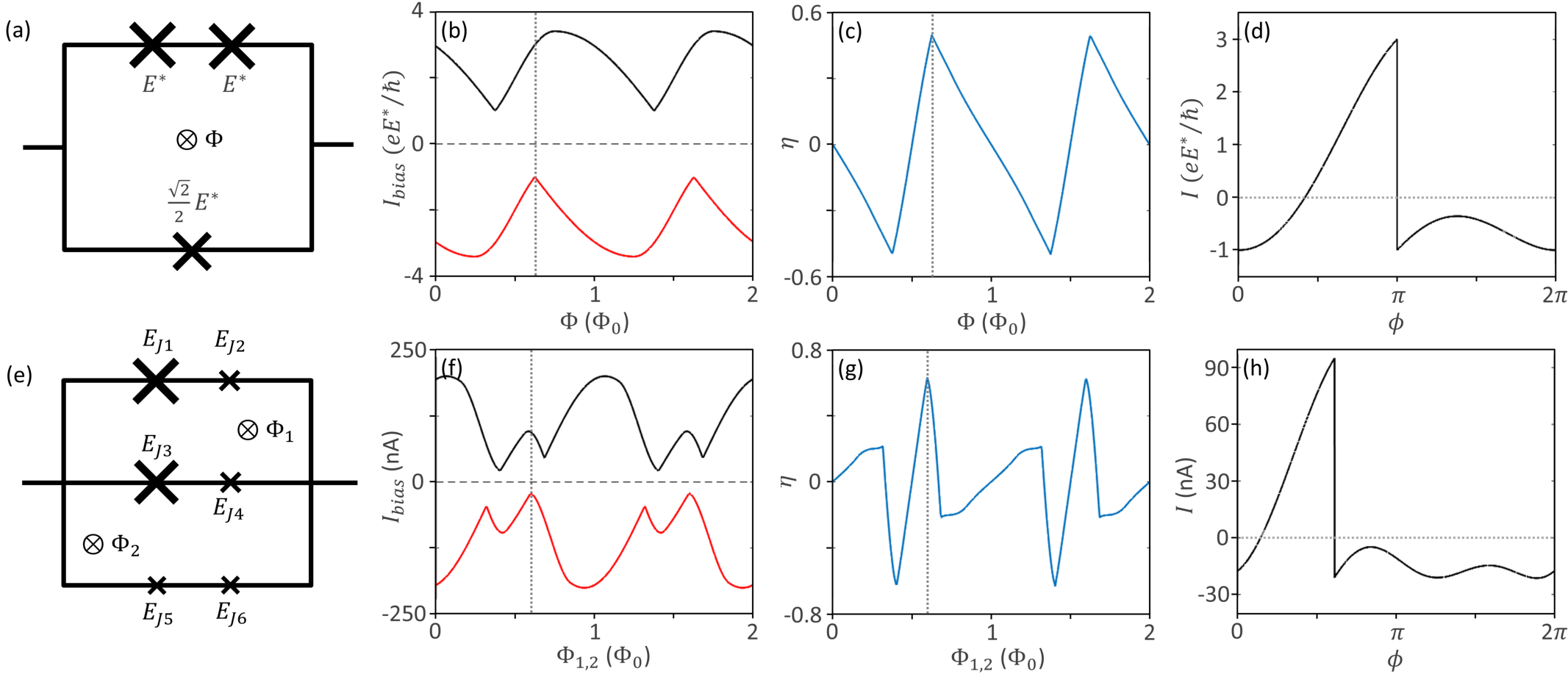}
    \caption{\label{fig:figS3}%
    Demonstration of single-loop and double-loop SQUID configurations optimizing $\eta$. (a) Schematic of device expected to have the CPR given by Eq.~(\ref{eq:eqS4}). (b) Modeled $I_c^+$ and $I_c^-$ as a function of external magnetic flux $\Phi$ for the single-loop device shown in Fig.~\ref{fig:figS3}(a). Calculated from device CPR given by Eq.~(\ref{eq:eqS4}). (c) $\eta$ as a function of external flux $\Phi$ calculated from the modeled switching currents in Fig.~\ref{fig:figS3}(b). (d) CPR $I(\phi)$ in Eq.~(\ref{eq:eqS4}) plotted for $\delta \phi=\frac{5\pi}{4}$. $I_c^+=\frac{3eE^*}{\hbar}$, $|I_c^-|=\frac{eE^*}{\hbar}$, and $\eta=50\%$. (e) Schematic of double-loop SQUID configuration expected to maximize $\eta$. The size of each `X' corresponds to the relative $E_J$ for each junction. (f) Modeled $I_c^+$ and $I_c^-$ as a function of external magnetic flux threading both loops, where $\Phi_1=\Phi_2$. Calculated from device CPR given by Eq.~(\ref{eq:eq3}) using $E_{Ji}$'s displayed in Table \ref{tab:tableS2}. (g) $\eta$ as a function of external flux ($\Phi_1=\Phi_2$) calculated from the modeled switching currents in Fig.~\ref{fig:figS3}(f). (h) Device CPR for double-loop SQUID configuration where $\eta$ is approximately maximized. $I_c^+=95.03$~nA, $|I_c^-|=21.36$~nA, and $\eta \sim 63.3\%$.
    }
\end{figure*}

A hypothetical device with this CPR could be realized in a SQUID where one branch has two JJs with equal $E_J$'s ($\tau_1=1$) and the other branch has a single JJ with a sinusoidal CPR ($\tau_2=0$), as is shown in Fig.~\ref{fig:figS3}(a). For a SQUID where $E_{J1}=E_{J2}=E^*$ on the top branch and $E_{J3}=\frac{\sqrt{2}}{2}E^*$ on the bottom branch, the total CPR of the SQUID would be given by: 

\begin{equation}
    \label{eq:eqS4}
    I_{SL,\eta_{max}}(\phi,\delta\phi) = \frac{eE^*}{\hbar}\left[\frac{\sin(\phi)}{\sqrt{1-\sin^2(\frac{\phi}{2})}} + \sqrt{2}\sin(\phi + \delta \phi)\right]
\end{equation}

This hypothetical device would be operable as a flux-tunable diode where $\eta$ could be set to any value between $-50\%$ and $50\%$ (Fig.~\ref{fig:figS3}(c)). Fig.~\ref{fig:figS3}(d) explicitly plots $I_{SL,\eta_{max}}(\phi,\delta\phi=\frac{5\pi}{4})$, where it can be seen that $I_c^+=3|I_c^-|$ and therefore, $\eta=50\%$.

A mathematical upper bound for the diode efficiency of $N_J$ interfering functions that aren't diodes themselves is given by $\eta^*=(N_J-1)/N_J$ \cite{Souto2022}. For the single loop case, we have two interfering branches, so $N_J=2$ and $\eta^*=(2-1)/2=50\%$, matching the diode efficiency of the function in Eq.~(\ref{eq:eqS4}). This mathematical upper bound $\eta^*$ was constructed initially with highly idealized functions (step functions and delta functions) in mind, so it is somewhat surprising that we can manage to construct a realizable device CPR where $\eta=\eta^*$.

\section{CPR OPTIMIZING $\eta$ IN A DOUBLE-LOOP SQUID}
\label{app:appE}
The CPR of a double-loop SQUID with two sinusoidal JJs on each branch is given by Eq.~(\ref{eq:eq3}). To determine a set of free parameters that yields a CPR with a maximized $\eta$, we perform another Monte Carlo optimization. To make this process more directly useful for experimental tuning, we allow $\{E_{Ji},\Phi_i\}$ to comprise our set of free parameters instead of $\{\tau_i,\sigma_i,\delta\phi_i\}$. Note that $\tau_i$ and $\sigma_i$ can be directly calculated from $E_{Ji,1}$ and $E_{Ji,2}$.

Choosing to randomize $E_J$'s instead of $\tau_i$'s and $\sigma_i$'s allows us to more easily impose constraints reflecting the lack of gate response from gates $V_1$ and $V_3$ in our device. As was discussed in the main text, we impose this constraint during device modeling by fixing $E_{J1}$ and $E_{J3}$ to the values displayed in Table \ref{tab:tableS2}. Additionally, since our device was flux-biased using an external magnetic field, and the areas of both loops were designed to be equal, we set $\Phi_1=\Phi_2$ for this optimization.

The volume of phase space that we need to optimize within increases exponentially with the number of loops, so it is helpful to reduce the number of free parameters used for MC optimization wherever possible. As was the case for the single-loop SQUID, we expect a maximized $\eta$ when one branch has maximal higher harmonic content. To achieve this, we impose the constraint $E_{J5}=E_{J6}$, reducing the number of free parameters by one.

With these constraints in mind, we follow a similar MC optimization procedure, where we randomize combinations of parameters $\{E_{J2}, E_{J4}, E_{J5} \textnormal{ (equal to } E_{J6}), \Phi_1 \textnormal{ (equal to } \Phi_2)\}$ from a uniform distribution and calculate $\eta$ numerically for a large number of iterations. The ranges of values we randomize our $E_J$'s to correspond to the range of energies we can set for each junction using gate tuning.

For the double-loop case, we are unable to converge to a single optimal device configuration where $\eta=\eta^*$, as we were able to do for the single-loop SQUID. This is due to both the constraints we imposed on the model to reflect our device's tunability and the fact that the parameter space we are optimizing in is significantly larger. This does not mean that a similar optimal solution where $\eta=\eta^*$ does not exist; however, it was not found in our search.

\begin{table}
\caption{\label{tab:tableS2} 
$E_{Ji}$'s from Monte Carlo optimization of a double-loop SQUID used to calculate modeled $I_c^+$ and $I_c^-$ oscillations in Fig.~\ref{fig:figS3}(f), along with the effective transparency of each branch in this model.
}
\centering
\begin{tabular*}{\linewidth}{@{\extracolsep{\fill}} @{\hspace{8mm}} c c c c c}
\toprule
$i$ &  $E_{Ji}$ (nA$\cdot\frac{\hbar}{2e}$) & $E_{Ji}$ (meV) & $\tau_{\mathrm{eff}}$ \\
\midrule
1 & 420 & 0.8633 & 0.514 \\
2 &  75 & 0.1542 &                       \\
\midrule
3 & 380 & 0.7811 & 0.551 \\
4 &  75 & 0.1542 &                       \\
\midrule
5 &  58 & 0.1192 & 1.000 \\
6 &  58 & 0.1192 &                       \\
\bottomrule
\end{tabular*}
\end{table}

Through our optimization procedure, we are still able to find a configuration where $\eta \sim 63.3\%$, which is only a few percent short of $\eta^*=\frac{2}{3}\sim66.7\%$ for the double-loop case ($N_J=3$). This optimal configuration is achieved when $\Phi_1=\Phi_2\sim 0.5972~\Phi_0$ and the $E_{Ji}$'s of all junctions are set to the values in Table \ref{tab:tableS2}. $I_{tot}(\phi)$ from Eq.~(\ref{eq:eq3}) is plotted for these parameter values in Fig.~\ref{fig:figS3}(h), demonstrating a CPR where $\eta \sim 63.3\%$. 

There were some challenges when attempting to use the gate mappings for each junction to quantitatively tune the $E_J$'s of the device to this optimal configuration, since these gate mappings would shift slightly over time. These shifts appeared to be around $10-20$~mV for each junction, so the gate mappings were still helpful in approaching the optimal voltage configuration of the device. However, this made it difficult to set all of the junction $E_J$'s to the exact values needed to observe the maximum possible diode efficiency achievable in our device of $\eta>60\%$. There is a range in parameter space around this optimal configuration where $|\eta|>50\%$ can be observed, and we can tune our device to a voltage configuration within this range (Section \ref{sec:sec3.1}).

\section{ADDITIONAL FIGURES}
\label{app:appF}

\begin{figure}
    \includegraphics[scale=0.5]{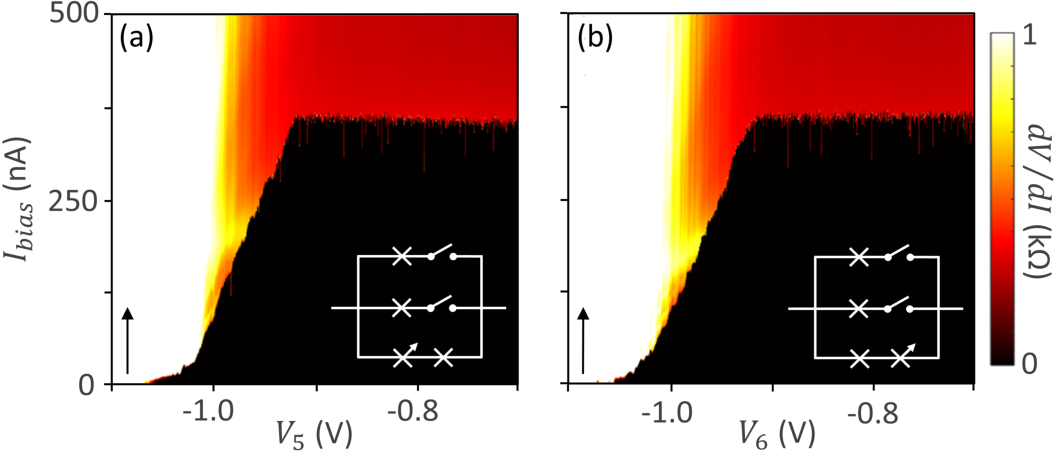}
    \caption{\label{fig:figS1}%
    (a) Differential resistance ($dV/dI$) plotted as a function of gate voltage $V_5$ and DC current bias $I_{bias}$. Gates $V_2$ and $V_4$ are set to voltages of $-1.1$V to pinch off the top and middle branches, respectively. Gate voltage $V_6$ is held at $0$~V. Black arrow denotes $I_{bias}$ sweep direction for the plotted data. (b) Same as Fig.~\ref{fig:figS1}(a), except sweeping gate voltage $V_6$ while keeping $V_5$ at $0$~V. Colorbar applies to both Fig.~\ref{fig:figS1}(a) and Fig.~\ref{fig:figS1}(b). 
    }
\end{figure}

\begin{figure}
    \includegraphics[scale=0.5]{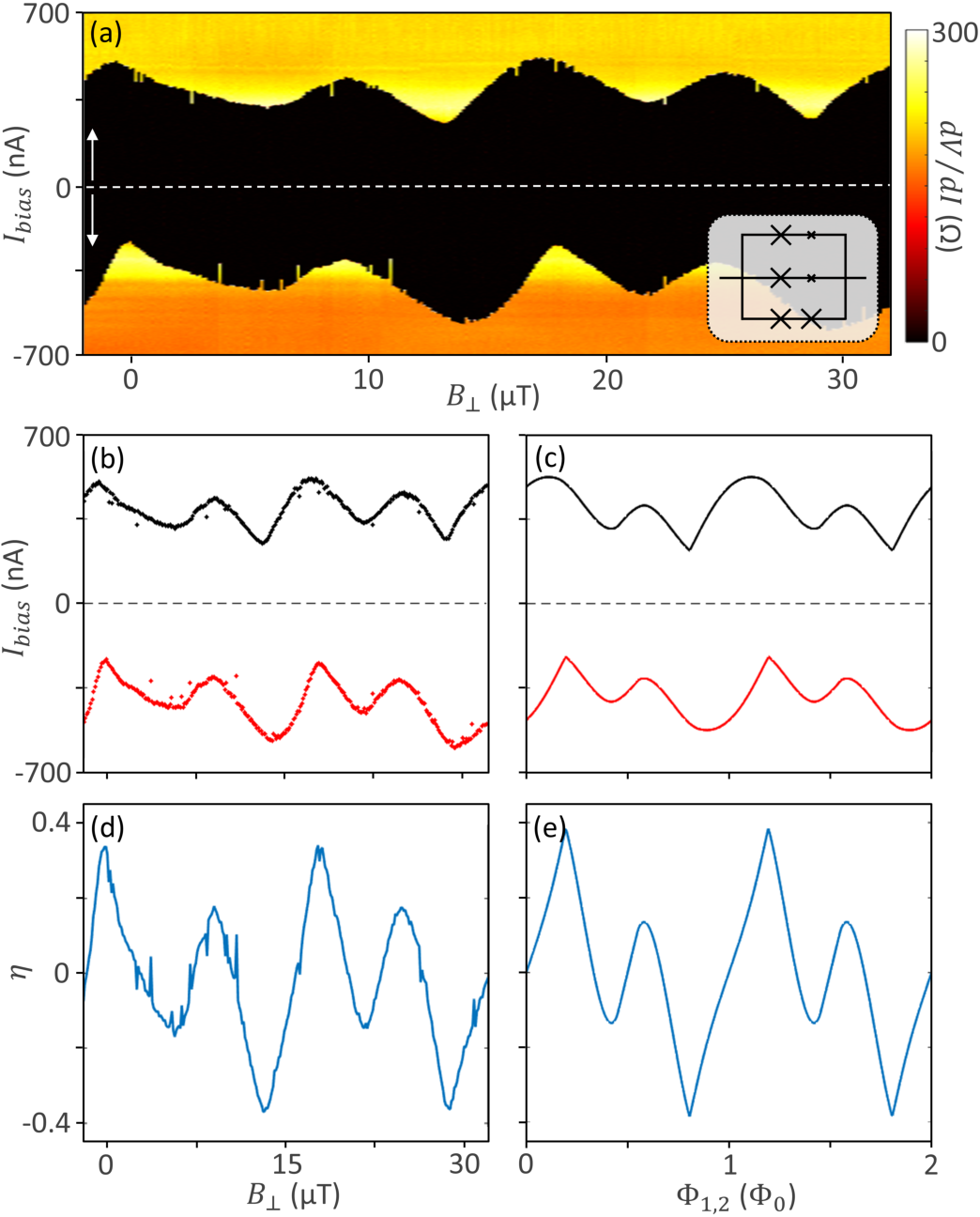}
    \caption{\label{fig:figS2}%
    (a) Differential resistance ($dV/dI$) as a function of $I_{bias}$ and $B_{\perp}$ with junction gates tuned to another configuration where significant asymmetry was observed (gate voltages applied to each junction are displayed in Table \ref{tab:tableS1}). (b) $I_c^+$ (black) and $I_c^-$ (red) as a function of $B_{\perp}$ extracted from the data plotted in Fig.~\ref{fig:figS2}(a) using a threshold resistance of $10\Omega$. (c) Modeled $I_c^+$ and $I_c^-$ as a function of external magnetic flux threading both loops, where $\Phi_1=\Phi_2$. Calculated from device CPR given by Eq.~(\ref{eq:eq3}) using $E_{Ji}$'s displayed in Table \ref{tab:tableS1}. (d) Measured diode efficiency $\eta$ as a function of $B_{\perp}$ calculated from the switching currents plotted in Fig.~\ref{fig:figS2}(b). (e) $\eta$ as a function of external flux ($\Phi_1=\Phi_2$) calculated from the modeled switching currents in Fig.~\ref{fig:figS2}(c).
    }
\end{figure}

\begin{table}
\caption{\label{tab:tableS1} 
Gate voltages applied to each JJ to generate data shown in Fig.~\ref{fig:figS2}(a). 
$E_{Ji}$'s used to calculate modeled $I_c^+$ and $I_c^-$ oscillations in Fig.~\ref{fig:figS2}(c) 
are also shown, along with the effective transparency of each branch in this model.
}
\centering
\begin{tabular*}{\linewidth}{@{\extracolsep{\fill}} c c c c c}
\toprule
$i$ & $V_i$ (V) & $E_{Ji}$ (nA$\cdot\frac{\hbar}{2e}$) & $E_{Ji}$ (meV) & $\tau_{\mathrm{eff}}$ \\
\midrule
1 & 0      & 420 & 0.8633 & 0.581 \\
2 & -1.028 &  90 & 0.1850 &                       \\
\midrule
3 & 0      & 380 & 0.7811 & 0.619 \\
4 & -1.019 &  90 & 0.1850 &                       \\
\midrule
5 & 0 & 360 & 0.7399 & 1.000 \\
6 & 0 & 360 & 0.7399 &                       \\
\bottomrule
\end{tabular*}
\end{table}

Fig.~\ref{fig:figS1} displays the gate voltage vs. $E_J$ maps for the junctions tuned by $V_5$ and $V_6$. This figure is generated using the same method used to generate Fig.~\ref{fig:fig2}(a) and Fig.~\ref{fig:fig2}(b). One of the junctions in the top and middle branches is fully depleted by applying a negative gate voltage so that supercurrent transport only occurs through the bottom branch. Then, according to Eq.~(\ref{eq:eq4}), the measured $I_c$ is proportional to the smaller of the two $E_J$'s in the bottom branch of the device. As was the case for the junctions controlled by $V_2$ and $V_4$, the change in $I_c$ that we measure is negligible until a gate voltage of approximately $-0.9$~V is applied to either junction whose response we are measuring. 

Fig.~\ref{fig:figS2} displays highly asymmetric SQUID oscillations measured in a second voltage configuration. To generate this figure, negative gate voltages are applied to $V_2$ and $V_4$ so that $E_{J1} \gg E_{J2}$, and $E_{J3} \gg E_{J4}$, and $E_{J5} \sim E_{J6}$. The difference between this configuration and the one we tune to in Section \ref{sec:sec3.1} is that here, we keep $V_5$ and $V_6$ set to $0$~V. As a result of this change, the CPR of the full device is affected significantly, and the measured oscillations in $I_c^+$ and $I_c^-$ take on an entirely different qualitative form.

Note that the measured $I_c$ when $V_{5,6}<-0.9$~V is approximately equivalent in Fig.~\ref{fig:figS1}(a) and Fig.~\ref{fig:figS1}(b). This does not necessarily mean that $E_{J5}$ and $E_{J6}$ are equivalent when no gate voltage is applied to either junction, since, according to Eq.~(\ref{eq:eq4}), we are only able to probe the smaller of these two energies with this measurement. Still, we assume that $E_{J5}=E_{J6}$ if $V_5=V_6=0$~V when modeling the device CPR in this voltage configuration, as is shown in Table \ref{tab:tableS1}.

This configuration also displays significant asymmetry that can be tuned using external flux to any $\eta$ between $-37\%$ and $34\%$, as is shown in Fig.~\ref{fig:figS2}(d). As was done in Section \ref{sec:sec3.1}, we model this data by numerically calculating the device CPR using Eq.~(\ref{eq:eq3}) with $E_{Ji}$'s for each junction now set to the values shown in Table \ref{tab:tableS1}. The resulting modeled oscillations in $I_c^+$ and $I_c^-$ shown in Fig.~\ref{fig:figS2}(c) again show explicit qualitative and quantitative agreement with our data.  

\nocite{*}
\bibliography{apssamp}

@PREAMBLE{
 "\providecommand{\noopsort}[1]{}" 
 # "\providecommand{\singleletter}[1]{#1}%" 
}

@article{Ambegaokar1969,
  title = {Voltage Due to Thermal Noise in the dc Josephson Effect},
  author = {Ambegaokar, Vinay and Halperin, B. I.},
  journal = {Phys. Rev. Lett.},
  volume = {22},
  issue = {25},
  pages = {1364--1366},
  numpages = {0},
  year = {1969},
  month = {Jun},
  publisher = {American Physical Society},
  doi = {10.1103/PhysRevLett.22.1364},
  url = {https://link.aps.org/doi/10.1103/PhysRevLett.22.1364}
}

@article{Marcus2024,
  title = {Voltage-Controlled Synthesis of Higher Harmonics in Hybrid Josephson Junction Circuits},
  author = {Banszerus, L. and Marshall, W. and Andersson, C. W. and Lindemann, T. and Manfra, M. J. and Marcus, C. M. and Vaitiek\ifmmode \dot{e}\else \.{e}\fi{}nas, S.},
  journal = {Phys. Rev. Lett.},
  volume = {133},
  issue = {18},
  pages = {186303},
  numpages = {6},
  year = {2024},
  month = {Oct},
  publisher = {American Physical Society},
  doi = {10.1103/PhysRevLett.133.186303},
  url = {https://link.aps.org/doi/10.1103/PhysRevLett.133.186303}
}

@article{Marcus2025,
  title = {Hybrid Josephson Rhombus: A Superconducting Element with Tailored Current-Phase Relation},
  author = {Banszerus, L. and Andersson, C. W. and Marshall, W. and Lindemann, T. and Manfra, M. J. and Marcus, C. M. and Vaitiek\ifmmode \dot{e}\else \.{e}\fi{}nas, S.},
  journal = {Phys. Rev. X},
  volume = {15},
  issue = {1},
  pages = {011021},
  numpages = {12},
  year = {2025},
  month = {Feb},
  publisher = {American Physical Society},
  doi = {10.1103/PhysRevX.15.011021},
  url = {https://link.aps.org/doi/10.1103/PhysRevX.15.011021}
}

@article{Schonenberger2023,
  title = {Gate-tunable Josephson diode in proximitized InAs supercurrent interferometers},
  author = {Ciaccia, Carlo and Haller, Roy and Drachmann, Asbj\o{}rn C. C. and Lindemann, Tyler and Manfra, Michael J. and Schrade, Constantin and Sch\"onenberger, Christian},
  journal = {Phys. Rev. Res.},
  volume = {5},
  issue = {3},
  pages = {033131},
  numpages = {12},
  year = {2023},
  month = {Aug},
  publisher = {American Physical Society},
  doi = {10.1103/PhysRevResearch.5.033131},
  url = {https://link.aps.org/doi/10.1103/PhysRevResearch.5.033131}
}

@article{Nichele2024,
	title = {Flux-{Tunable} {Josephson} {Diode} {Effect} in a {Hybrid} {Four}-{Terminal} {Josephson} {Junction}},
	volume = {18},
	issn = {1936-0851, 1936-086X},
	url = {http://arxiv.org/abs/2312.04415},
	doi = {10.1021/acsnano.4c01642},
	number = {12},
	urldate = {2024-07-16},
	journal = {ACS Nano},
	author = {Coraiola, M. and Svetogorov, A. E. and Haxell, D. Z. and Sabonis, D. and Hinderling, M. and Kate, S. C. ten and Cheah, E. and Krizek, F. and Schott, R. and Wegscheider, W. and Cuevas, J. C. and Belzig, W. and Nichele, F.},
	month = mar,
	year = {2024},
	note = {arXiv:2312.04415 [cond-mat]},
	pages = {9221--9231},
}

@article{Greco2024,
	title = {Double loop dc-{SQUID} as a tunable {Josephson} diode},
	volume = {125},
	issn = {0003-6951},
	url = {https://doi.org/10.1063/5.0211021},
	doi = {10.1063/5.0211021},
	number = {7},
	journal = {Applied Physics Letters},
	author = {Greco, A. and Pichard, Q. and Strambini, E. and Giazotto, F.},
	month = aug,
	year = {2024},
	pages = {072601},
}

@article{ABS1991,
  title = {Universal limit of critical-current fluctuations in mesoscopic Josephson junctions},
  author = {Beenakker, C. W. J.},
  journal = {Phys. Rev. Lett.},
  volume = {67},
  issue = {27},
  pages = {3836--3839},
  numpages = {0},
  year = {1991},
  month = {Dec},
  publisher = {American Physical Society},
  doi = {10.1103/PhysRevLett.67.3836},
  url = {https://link.aps.org/doi/10.1103/PhysRevLett.67.3836}
}

@Article{Bozkurt2023,
	title={{Double-Fourier engineering of Josephson energy-phase relationships  applied to diodes}},
	author={A. Mert Bozkurt and Jasper Brookman and Valla Fatemi and Anton R. Akhmerov},
	journal={SciPost Phys.},
	volume={15},
	pages={204},
	year={2023},
	publisher={SciPost},
	doi={10.21468/SciPostPhys.15.5.204},
	url={https://scipost.org/10.21468/SciPostPhys.15.5.204},
}

@article{Souto2022,
  title = {Josephson Diode Effect in Supercurrent Interferometers},
  author = {Souto, Rub\'en Seoane and Leijnse, Martin and Schrade, Constantin},
  journal = {Phys. Rev. Lett.},
  volume = {129},
  issue = {26},
  pages = {267702},
  numpages = {6},
  year = {2022},
  month = {Dec},
  publisher = {American Physical Society},
  doi = {10.1103/PhysRevLett.129.267702},
  url = {https://link.aps.org/doi/10.1103/PhysRevLett.129.267702}
}

@article{Barash2018,
  title = {Proximity-reduced range of internal phase differences in double Josephson junctions with closely spaced interfaces},
  author = {Barash, Yu. S.},
  journal = {Phys. Rev. B},
  volume = {97},
  issue = {22},
  pages = {224509},
  numpages = {7},
  year = {2018},
  month = {Jun},
  publisher = {American Physical Society},
  doi = {10.1103/PhysRevB.97.224509},
  url = {https://link.aps.org/doi/10.1103/PhysRevB.97.224509}
}

@article{Nichele2020,
  title = {Relating Andreev Bound States and Supercurrents in Hybrid Josephson Junctions},
  author = {Nichele, F. and Portol\'es, E. and Fornieri, A. and Whiticar, A. M. and Drachmann, A. C. C. and Gronin, S. and Wang, T. and Gardner, G. C. and Thomas, C. and Hatke, A. T. and Manfra, M. J. and Marcus, C. M.},
  journal = {Phys. Rev. Lett.},
  volume = {124},
  issue = {22},
  pages = {226801},
  numpages = {6},
  year = {2020},
  month = {Jun},
  publisher = {American Physical Society},
  doi = {10.1103/PhysRevLett.124.226801},
  url = {https://link.aps.org/doi/10.1103/PhysRevLett.124.226801}
}

@article{Leblanc2024,
   title={From nonreciprocal to charge-4e supercurrent in Ge-based Josephson devices with tunable harmonic content},
   volume={6},
   ISSN={2643-1564},
   url={http://dx.doi.org/10.1103/PhysRevResearch.6.033281},
   DOI={10.1103/physrevresearch.6.033281},
   number={3},
   journal={Physical Review Research},
   publisher={American Physical Society (APS)},
   author={Leblanc, Axel and Tangchingchai, Chotivut and Momtaz, Zahra Sadre and Kiyooka, Elyjah and Hartmann, Jean-Michel and Fernandez-Bada, Gonzalo Troncoso and Scherübl, Zoltán and Brun, Boris and Schmitt, Vivien and Zihlmann, Simon and others},
   others = {Maurand, Romain and Dumur, Étienne and De Franceschi, Silvano and Lefloch, François},
   year={2024},
   month=sep }

@article{Josephson1962,
title = {Possible new effects in superconductive tunnelling},
journal = {Physics Letters},
volume = {1},
number = {7},
pages = {251-253},
year = {1962},
issn = {0031-9163},
doi = {https://doi.org/10.1016/0031-9163(62)91369-0},
url = {https://www.sciencedirect.com/science/article/pii/0031916362913690},
author = {B.D. Josephson}
}

@article{DellaRocca2007,
  title = {Measurement of the Current-Phase Relation of Superconducting Atomic Contacts},
  author = {Della Rocca, M. L. and Chauvin, M. and Huard, B. and Pothier, H. and Esteve, D. and Urbina, C.},
  journal = {Phys. Rev. Lett.},
  volume = {99},
  issue = {12},
  pages = {127005},
  numpages = {4},
  year = {2007},
  month = {Sep},
  publisher = {American Physical Society},
  doi = {10.1103/PhysRevLett.99.127005},
  url = {https://link.aps.org/doi/10.1103/PhysRevLett.99.127005}
}

@article{Likharev1979,
  title = {Superconducting weak links},
  author = {Likharev, K. K.},
  journal = {Rev. Mod. Phys.},
  volume = {51},
  issue = {1},
  pages = {101--159},
  numpages = {0},
  year = {1979},
  month = {Jan},
  publisher = {American Physical Society},
  doi = {10.1103/RevModPhys.51.101},
  url = {https://link.aps.org/doi/10.1103/RevModPhys.51.101}
}

@article{TransmonChargeDisp2020,
  title = {Suppressed Charge Dispersion via Resonant Tunneling in a Single-Channel Transmon},
  author = {Kringh\o{}j, A. and van Heck, B. and Larsen, T. W. and Erlandsson, O. and Sabonis, D. and Krogstrup, P. and Casparis, L. and Petersson, K. D. and Marcus, C. M.},
  journal = {Phys. Rev. Lett.},
  volume = {124},
  issue = {24},
  pages = {246803},
  numpages = {6},
  year = {2020},
  month = {Jun},
  publisher = {American Physical Society},
  doi = {10.1103/PhysRevLett.124.246803},
  url = {https://link.aps.org/doi/10.1103/PhysRevLett.124.246803}
}

@article{SCIslandChargeDisp2020,
  title = {Observation of Vanishing Charge Dispersion of a Nearly Open Superconducting Island},
  author = {Bargerbos, Arno and Uilhoorn, Willemijn and Yang, Chung-Kai and Krogstrup, Peter and Kouwenhoven, Leo P. and de Lange, Gijs and van Heck, Bernard and Kou, Angela},
  journal = {Phys. Rev. Lett.},
  volume = {124},
  issue = {24},
  pages = {246802},
  numpages = {7},
  year = {2020},
  month = {Jun},
  publisher = {American Physical Society},
  doi = {10.1103/PhysRevLett.124.246802},
  url = {https://link.aps.org/doi/10.1103/PhysRevLett.124.246802}
}

@article{Willsch2024,
	title = {Observation of {Josephson} harmonics in tunnel junctions},
	volume = {20},
	issn = {1745-2481},
	url = {https://doi.org/10.1038/s41567-024-02400-8},
	doi = {10.1038/s41567-024-02400-8},
	number = {5},
	journal = {Nature Physics},
	author = {Willsch, Dennis and Rieger, Dennis and Winkel, Patrick and Willsch, Madita and Dickel, Christian and Krause, Jonas and Ando, Yoichi and Lescanne, Raphaël and Leghtas, Zaki and Bronn, Nicholas T. and Deb, Pratiti and others},
    other = {and Lanes, Olivia and Minev, Zlatko K. and Dennig, Benedikt and Geisert, Simon and Günzler, Simon and Ihssen, Sören and Paluch, Patrick and Reisinger, Thomas and Hanna, Roudy and Bae, Jin Hee and Schüffelgen, Peter and Grützmacher, Detlev and Buimaga-Iarinca, Luiza and Morari, Cristian and Wernsdorfer, Wolfgang and DiVincenzo, David P. and Michielsen, Kristel and Catelani, Gianluigi and Pop, Ioan M.},
	month = may,
	year = {2024},
	pages = {815--821},
}

@article{Baumgartner2022,
	title = {Supercurrent rectification and magnetochiral effects in symmetric {Josephson} junctions},
	volume = {17},
	issn = {1748-3395},
	url = {https://doi.org/10.1038/s41565-021-01009-9},
	doi = {10.1038/s41565-021-01009-9},
	number = {1},
	journal = {Nature Nanotechnology},
	author = {Baumgartner, Christian and Fuchs, Lorenz and Costa, Andreas and Reinhardt, Simon and Gronin, Sergei and Gardner, Geoffrey C. and Lindemann, Tyler and Manfra, Michael J. and Faria Junior, Paulo E. and Kochan, Denis and Fabian, Jaroslav and Paradiso, Nicola and Strunk, Christoph},
	month = jan,
	year = {2022},
	pages = {39--44},
}

@article{Ciaccia2024,
	title = {Charge-4e supercurrent in a two-dimensional {InAs}-{Al} superconductor-semiconductor heterostructure},
	volume = {7},
	issn = {2399-3650},
	url = {https://doi.org/10.1038/s42005-024-01531-x},
	doi = {10.1038/s42005-024-01531-x},
	number = {1},
	journal = {Communications Physics},
	author = {Ciaccia, Carlo and Haller, Roy and Drachmann, Asbjørn C. C. and Lindemann, Tyler and Manfra, Michael J. and Schrade, Constantin and Schönenberger, Christian},
	month = jan,
	year = {2024},
	pages = {41},
}

@article{JJReview2004,
  title = {The current-phase relation in Josephson junctions},
  author = {Golubov, A. A. and Kupriyanov, M. Yu. and Il'ichev, E.},
  journal = {Rev. Mod. Phys.},
  volume = {76},
  issue = {2},
  pages = {411--469},
  numpages = {0},
  year = {2004},
  month = {Apr},
  publisher = {American Physical Society},
  doi = {10.1103/RevModPhys.76.411},
  url = {https://link.aps.org/doi/10.1103/RevModPhys.76.411}
}

@article{JJArrayQubit2022,
  title = {Protected Hybrid Superconducting Qubit in an Array of Gate-Tunable Josephson Interferometers},
  author = {Schrade, Constantin and Marcus, Charles M. and Gyenis, Andr\'as},
  journal = {PRX Quantum},
  volume = {3},
  issue = {3},
  pages = {030303},
  numpages = {15},
  year = {2022},
  month = {Jul},
  publisher = {American Physical Society},
  doi = {10.1103/PRXQuantum.3.030303},
  url = {https://link.aps.org/doi/10.1103/PRXQuantum.3.030303}
}

@article{Smith2020,
	title = {Superconducting circuit protected by two-{Cooper}-pair tunneling},
	volume = {6},
	issn = {2056-6387},
	url = {https://doi.org/10.1038/s41534-019-0231-2},
	doi = {10.1038/s41534-019-0231-2},
	number = {1},
	journal = {npj Quantum Information},
	author = {Smith, W. C. and Kou, A. and Xiao, X. and Vool, U. and Devoret, M. H.},
	month = jan,
	year = {2020},
	pages = {8},
}

@article{tinkham1996,
	title = {Introduction to {Superconductivity}},
	volume = {49},
	issn = {0031-9228},
	url = {https://doi.org/10.1063/1.2807811},
	doi = {10.1063/1.2807811},
	number = {10},
	journal = {Physics Today},
	author = {Tinkham, Michael and Emery, Victor},
	month = oct,
	year = {1996},
	pages = {198--198},
}

@article{Ando2020,
	title = {Observation of superconducting diode effect},
	volume = {584},
	issn = {1476-4687},
	url = {https://doi.org/10.1038/s41586-020-2590-4},
	doi = {10.1038/s41586-020-2590-4},
	number = {7821},
	journal = {Nature},
	author = {Ando, Fuyuki and Miyasaka, Yuta and Li, Tian and Ishizuka, Jun and Arakawa, Tomonori and Shiota, Yoichi and Moriyama, Takahiro and Yanase, Youichi and Ono, Teruo},
	month = aug,
	year = {2020},
	pages = {373--376},
}

@article{ThinFilmTheory2022,
  title = {Intrinsic Superconducting Diode Effect},
  author = {Daido, Akito and Ikeda, Yuhei and Yanase, Youichi},
  journal = {Phys. Rev. Lett.},
  volume = {128},
  issue = {3},
  pages = {037001},
  numpages = {6},
  year = {2022},
  month = {Jan},
  publisher = {American Physical Society},
  doi = {10.1103/PhysRevLett.128.037001},
  url = {https://link.aps.org/doi/10.1103/PhysRevLett.128.037001}
}

@article{Hou2023,
  title = {Ubiquitous Superconducting Diode Effect in Superconductor Thin Films},
  author = {Hou, Yasen and Nichele, Fabrizio and Chi, Hang and Lodesani, Alessandro and Wu, Yingying and Ritter, Markus F. and Haxell, Daniel Z. and Davydova, Margarita and Ili\ifmmode \acute{c}\else \'{c}\fi{}, Stefan and Glezakou-Elbert, Ourania and others},
  others = {Varambally, Amith and Bergeret, F. Sebastian and Kamra, Akashdeep and Fu, Liang and Lee, Patrick A. and Moodera, Jagadeesh S.},
  journal = {Phys. Rev. Lett.},
  volume = {131},
  issue = {2},
  pages = {027001},
  numpages = {6},
  year = {2023},
  month = {Jul},
  publisher = {American Physical Society},
  doi = {10.1103/PhysRevLett.131.027001},
  url = {https://link.aps.org/doi/10.1103/PhysRevLett.131.027001}
}

@article{RashbaTheory2022,
  title = {Theory of the Supercurrent Diode Effect in Rashba Superconductors with Arbitrary Disorder},
  author = {Ili\ifmmode \acute{c}\else \'{c}\fi{}, S. and Bergeret, F. S.},
  journal = {Phys. Rev. Lett.},
  volume = {128},
  issue = {17},
  pages = {177001},
  numpages = {6},
  year = {2022},
  month = {Apr},
  publisher = {American Physical Society},
  doi = {10.1103/PhysRevLett.128.177001},
  url = {https://link.aps.org/doi/10.1103/PhysRevLett.128.177001}
}

@article{Graphene2022,
	title = {Zero-field superconducting diode effect in small-twist-angle trilayer graphene},
	volume = {18},
	issn = {1745-2481},
	url = {https://doi.org/10.1038/s41567-022-01700-1},
	doi = {10.1038/s41567-022-01700-1},
	number = {10},
	journal = {Nature Physics},
	author = {Lin, Jiang-Xiazi and Siriviboon, Phum and Scammell, Harley D. and Liu, Song and Rhodes, Daniel and Watanabe, K. and Taniguchi, T. and Hone, James and Scheurer, Mathias S. and Li, J.I.A.},
	month = oct,
	year = {2022},
	pages = {1221--1227},
}

@article{GrapheneTheory2022,
doi = {10.1088/2053-1583/ac5b16},
url = {https://dx.doi.org/10.1088/2053-1583/ac5b16},
year = {2022},
month = {mar},
publisher = {IOP Publishing},
volume = {9},
number = {2},
pages = {025027},
author = {Scammell, Harley D and Li, J I A and Scheurer, Mathias S},
title = {Theory of zero-field superconducting diode effect in twisted trilayer graphene},
journal = {2D Materials},
}

@article{vdW2022,
	title = {The field-free {Josephson} diode in a van der {Waals} heterostructure},
	volume = {604},
	issn = {1476-4687},
	url = {https://doi.org/10.1038/s41586-022-04504-8},
	doi = {10.1038/s41586-022-04504-8},
	number = {7907},
	journal = {Nature},
	author = {Wu, Heng and Wang, Yaojia and Xu, Yuanfeng and Sivakumar, Pranava K. and Pasco, Chris and Filippozzi, Ulderico and Parkin, Stuart S. P. and Zeng, Yu-Jia and McQueen, Tyrel and Ali, Mazhar N.},
	month = apr,
	year = {2022},
	pages = {653--656},
}

@article{vdWTheory2022,
  title = {General Theory of Josephson Diodes},
  author = {Zhang, Yi and Gu, Yuhao and Li, Pengfei and Hu, Jiangping and Jiang, Kun},
  journal = {Phys. Rev. X},
  volume = {12},
  issue = {4},
  pages = {041013},
  numpages = {11},
  year = {2022},
  month = {Nov},
  publisher = {American Physical Society},
  doi = {10.1103/PhysRevX.12.041013},
  url = {https://link.aps.org/doi/10.1103/PhysRevX.12.041013}
}

@article{TopoJJ2018,
  title = {Asymmetric Josephson effect in inversion symmetry breaking topological materials},
  author = {Chen, Chui-Zhen and He, James Jun and Ali, Mazhar N. and Lee, Gil-Ho and Fong, Kin Chung and Law, K. T.},
  journal = {Phys. Rev. B},
  volume = {98},
  issue = {7},
  pages = {075430},
  numpages = {5},
  year = {2018},
  month = {Aug},
  publisher = {American Physical Society},
  doi = {10.1103/PhysRevB.98.075430},
  url = {https://link.aps.org/doi/10.1103/PhysRevB.98.075430}
}

@article{TopoJJ2020,
	title = {One-{Dimensional} {Edge} {Transport} in {Few}-{Layer} {WTe2}},
	volume = {20},
	issn = {1530-6984},
	url = {https://doi.org/10.1021/acs.nanolett.0c00658},
	doi = {10.1021/acs.nanolett.0c00658},
	number = {6},
	journal = {Nano Letters},
	author = {Kononov, Artem and Abulizi, Gulibusitan and Qu, Kejian and Yan, Jiaqiang and Mandrus, David and Watanabe, Kenji and Taniguchi, Takashi and Schönenberger, Christian},
	month = jun,
	year = {2020},
	note = {Publisher: American Chemical Society},
	pages = {4228--4233},
}

@article{FiniteA2022,
	title = {Josephson diode effect from {Cooper} pair momentum in a topological semimetal},
	volume = {18},
	issn = {1745-2481},
	url = {https://doi.org/10.1038/s41567-022-01699-5},
	doi = {10.1038/s41567-022-01699-5},
	number = {10},
	journal = {Nature Physics},
	author = {Pal, Banabir and Chakraborty, Anirban and Sivakumar, Pranava K. and Davydova, Margarita and Gopi, Ajesh K. and Pandeya, Avanindra K. and Krieger, Jonas A. and Zhang, Yang and Date, Mihir and Ju, Sailong and Yuan, Noah and Schröter, Niels B. M. and Fu, Liang and Parkin, Stuart S. P.},
	month = oct,
	year = {2022},
	pages = {1228--1233},
}

@article{FiniteB2022,
author = {Margarita Davydova  and Saranesh Prembabu  and Liang Fu },
title = {Universal Josephson diode effect},
journal = {Science Advances},
volume = {8},
number = {23},
pages = {eabo0309},
year = {2022},
doi = {10.1126/sciadv.abo0309},
URL = {https://www.science.org/doi/abs/10.1126/sciadv.abo0309},
eprint = {https://www.science.org/doi/pdf/10.1126/sciadv.abo0309},
}

@article{FiniteC2022,
author = {Noah F. Q. Yuan  and Liang Fu },
title = {Supercurrent diode effect and finite-momentum superconductors},
journal = {Proceedings of the National Academy of Sciences},
volume = {119},
number = {15},
pages = {e2119548119},
year = {2022},
doi = {10.1073/pnas.2119548119},
URL = {https://www.pnas.org/doi/abs/10.1073/pnas.2119548119},
eprint = {https://www.pnas.org/doi/pdf/10.1073/pnas.2119548119},
}

@article{FluxFocus2021,
  title = {Theoretical model for parallel SQUID arrays with fluxoid focusing},
  author = {M\"uller, K.-H. and Mitchell, E. E.},
  journal = {Phys. Rev. B},
  volume = {103},
  issue = {5},
  pages = {054509},
  numpages = {14},
  year = {2021},
  month = {Feb},
  publisher = {American Physical Society},
  doi = {10.1103/PhysRevB.103.054509},
  url = {https://link.aps.org/doi/10.1103/PhysRevB.103.054509}
}

@article{Tarucha2023,
	title = {Josephson diode effect derived from short-range coherent coupling},
	volume = {19},
	issn = {1745-2481},
	url = {https://doi.org/10.1038/s41567-023-02144-x},
	doi = {10.1038/s41567-023-02144-x},
	number = {11},
	journal = {Nature Physics},
	author = {Matsuo, Sadashige and Imoto, Takaya and Yokoyama, Tomohiro and Sato, Yosuke and Lindemann, Tyler and Gronin, Sergei and Gardner, Geoffrey C. and Manfra, Michael J. and Tarucha, Seigo},
	month = nov,
	year = {2023},
	pages = {1636--1641},
}

@article{TaruchaAnomalous,
author = {Sadashige Matsuo  and Takaya Imoto  and Tomohiro Yokoyama  and Yosuke Sato  and Tyler Lindemann  and Sergei Gronin  and Geoffrey C. Gardner  and Michael J. Manfra  and Seigo Tarucha },
title = {Phase engineering of anomalous Josephson effect derived from Andreev molecules},
journal = {Science Advances},
volume = {9},
number = {50},
pages = {eadj3698},
year = {2023},
doi = {10.1126/sciadv.adj3698},
URL = {https://www.science.org/doi/abs/10.1126/sciadv.adj3698},
eprint = {https://www.science.org/doi/pdf/10.1126/sciadv.adj3698},
}

@article{Regensberg2025,
  title = {Unconventional Josephson supercurrent diode effect induced by chiral spin-orbit coupling},
  author = {Costa, Andreas and Kanehira, Osamu and Matsueda, Hiroaki and Fabian, Jaroslav},
  journal = {Phys. Rev. B},
  volume = {111},
  issue = {14},
  pages = {L140506},
  numpages = {9},
  year = {2025},
  month = {Apr},
  publisher = {American Physical Society},
  doi = {10.1103/PhysRevB.111.L140506},
  url = {https://link.aps.org/doi/10.1103/PhysRevB.111.L140506}
}

@article{Regensberg2024,
	title = {Link between supercurrent diode and anomalous {Josephson} effect revealed by gate-controlled interferometry},
	volume = {15},
	issn = {2041-1723},
	url = {https://doi.org/10.1038/s41467-024-48741-z},
	doi = {10.1038/s41467-024-48741-z},
	number = {1},
	journal = {Nature Communications},
	author = {Reinhardt, S. and Ascherl, T. and Costa, A. and Berger, J. and Gronin, S. and Gardner, G. C. and Lindemann, T. and Manfra, M. J. and Fabian, J. and Kochan, D. and Strunk, C. and Paradiso, N.},
	month = may,
	year = {2024},
	pages = {4413},
}

@article{Regensberg2023,
	title = {Sign reversal of the {Josephson} inductance magnetochiral anisotropy and 0–π-like transitions in supercurrent diodes},
	volume = {18},
	issn = {1748-3395},
	url = {https://doi.org/10.1038/s41565-023-01451-x},
	doi = {10.1038/s41565-023-01451-x},
	number = {11},
	journal = {Nature Nanotechnology},
	author = {Costa, A. and Baumgartner, C. and Reinhardt, S. and Berger, J. and Gronin, S. and Gardner, G. C. and Lindemann, T. and Manfra, M. J. and Fabian, J. and Kochan, D. and Paradiso, N. and Strunk, C.},
	month = nov,
	year = {2023},
	pages = {1266--1272},
}

@misc{SDEReview2025,
      title={Theories of Superconducting Diode Effects}, 
      author={Daniel Shaffer and Alex Levchenko},
      year={2025},
      eprint={2510.25864},
      archivePrefix={arXiv},
      primaryClass={cond-mat.supr-con},
      url={https://arxiv.org/abs/2510.25864}, 
}

\end{document}